\documentstyle[12pt]{article}

\input epsf

   \def\unlock{\catcode`@=11}

   \unlock

   \def\gsim{\mathrel{\mathpalette\@versim>}}

   \def\@versim#1#2{\vcenter{\offinterlineskip
        \ialign{$\m@th#1\hfil##\hfil$\crcr#2\crcr\sim\crcr } }}

\begin{document}

\begin{titlepage}

%% Begin SCIPP titlepage header with tree

%\frontpagetrue

\let\picnaturalsize=N
\def\picsize{1.0in}
\def\picfilename{scipp_tree.ps}

%If you do not have the picture file add:
%\let\nopictures=Y
%to the beginning of the file.

\ifx\nopictures Y\else
{\hbox to\hsize{\hbox{\ifx\picnaturalsize N\epsfxsize \picsize\fi
{\epsfbox{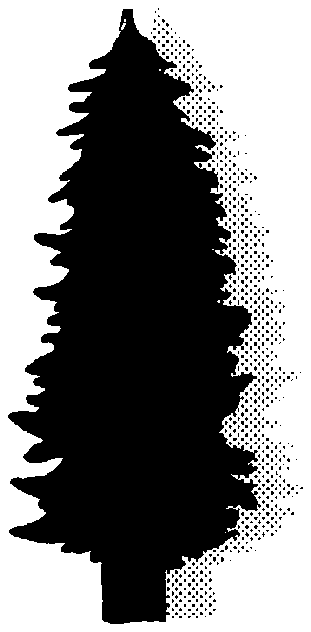}}}\hspace*{\fill}

% preprint and date go inside hboxes:
\parbox[b]{3.4cm}{SCIPP 95/40 \\ August 1995 \\ hep-ph/9509300\\
\vspace*{2.5cm}}

}}\fi

%% End of SCIPP titlepage header with tree

%\vspace*{1cm}

\begin{center}
\large\bf
Top Quark Physics at the NLC\footnote{Invited talk presented
at the Workshop on Physics of the Top Quark, Ames, Iowa, May 25 \& 26,
1995}
\end{center}

\vspace*{0.5cm}

\begin{center}
Carl R. Schmidt \footnote{Supported in part by the U.S.
Department of Energy.} \\
Santa Cruz Institute for Particle Physics\\
University of California, Santa Cruz, CA 95064, USA\\
\end{center}

\vspace*{0.5cm}

\begin{center}
\bf Abstract
\end{center}

\noindent
A high energy $e^+e^-$ linear collider (NLC) is an excellent tool
for studying the properties of the top quark.  In this talk I review
some of the theory of top quark production and decay in $e^+e^-$
collisions both at threshold and in the continuum.  I also report on the
results of phenomenological analyses of $t\bar t$ production at the NLC.
\end{titlepage}
\baselineskip=0.8cm

\section{Introduction}

A high energy electron-positron collider, which I will generically
refer to as the
Next Linear Collider or NLC, is an ideal tool for studying the
properties of the top quark.  The event environment in $e^+e^-$
collisions is clean so that precision measurements are possible.
The luminosity, which is expected to be on the order
of 50 fb$^{-1}/$yr, is sufficient to provide a yearly sample of a
few times $10^4$ top pairs.  In addition the high degree of electron
polarization attainable at the NLC will be very useful for probing the top
quark couplings to the photon and $Z$ boson.  Although I will not discuss it
here, there is also the possibility for a high-energy photon collider mode
using back-scattered lasers, which adds significantly to the versatility
of the machine.  There have been a number of studies of top quark physics
at the NLC\cite{NLCworkshops}.  In this talk I will discuss some of the
highlights, and I will report on the results of a few selected
phenomenological studies.

The CDF and D0 collaborations at Fermilab have obtained mass
values for the top quark of $176\pm8\pm10$ and $199^{+19}_{-21}\pm22$ GeV,
respectively\cite{CDFD0}.  This large mass indicates that the top quark
feels the strongest coupling to the symmetry breaking sector of any of the
observed particles.
Thus, the top quark interactions are an obvious place to look for hints
into the dynamics
of symmetry breaking.  More prosaically, the large mass has an important
impact on top quark phenomenology.  For large $m_t$, the top width grows
as
\begin{equation}
\Gamma_t\ \sim\ 1.7\ {\rm GeV}\ \biggl[{m_t\over175\ {\rm GeV}}\biggr]^3
\end{equation}
This implies that the time required for a top quark to decay is almost
always less than the time for hadronization to occur.  The decay time is
also significantly less
than the time it takes for a top quark  to depolarize.  Therefore, to a
good approximation,
the top quark can be treated as a free quark which transfers its
polarization to its decay products with little effect from hadronization.

\section{The Top at Threshold}

The large top width is important at threshold because
one of the quarks will typically decay before the $t\bar t$ pair
can form a bound state.  Thus, unlike the case for $c\bar c$ and
$b\bar b$, the $t\bar t$ threshold will not be dominated by large resonances.
However, for this very reason, it is possible to perform an accurate
calculation of the $t\bar t$ threshold cross section in QCD perturbation
theory.  At the $n$th order in
perturbation theory one finds a correction to the photon-$t\bar t$
vertex of order $(\alpha_s/v)^n$, where $v$ is the top quark velocity.
The infinite sum of these corrections which diverge as $v\rightarrow0$
can be
written as the solution to a Schr\"odinger Green's function equation
\begin{equation}
\Bigl[-{\nabla^2\over m_t}+V(\vec r)-E\Bigr]G_E(\vec r)\ =\ \delta^3(\vec r)
\end{equation}
with $E=\sqrt{s}-2m_t$ and with
the QCD potential $V(\vec r) = -(4/3)\alpha_s/r$ at short
distances.
The total cross section to $t\bar t$ pairs is then given in terms of the
Green's function by $\sigma\sim {\rm Im}G_E(0)$.

For the bottom and charm quarks, the singularity at small velocities is
cured by the QCD confining potential at long distances.  For the top,
however, Fadin and Khoze\cite{FadinKhoze} showed that it was possible to
include the width effects
by replacing $E\rightarrow E+i\Gamma_t$.  The singularity is then cut off by
$\Gamma_t$ with little dependence on the long distance potential.
The $t\bar t$ threshold cross section can be unambiguously computed as a
function of $\alpha_s$, $\Gamma_t$, and the top mass $m_t$.

\begin{figure}
\vskip10pt
\vskip-2.7cm
\epsfysize=11cm
\centerline{\hskip.5cm\epsffile{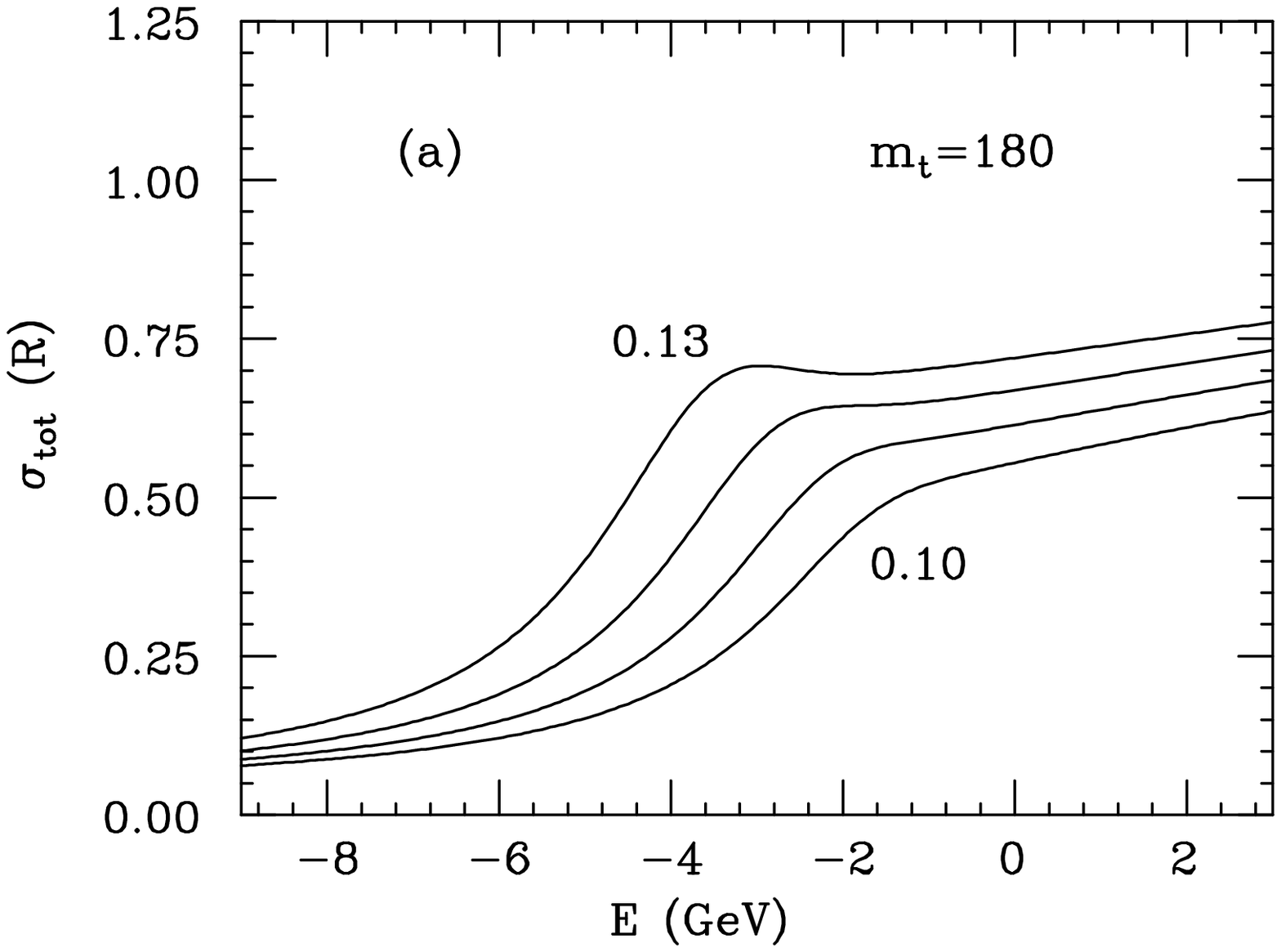}\hskip-1cm
             \epsfysize=11cm\epsffile{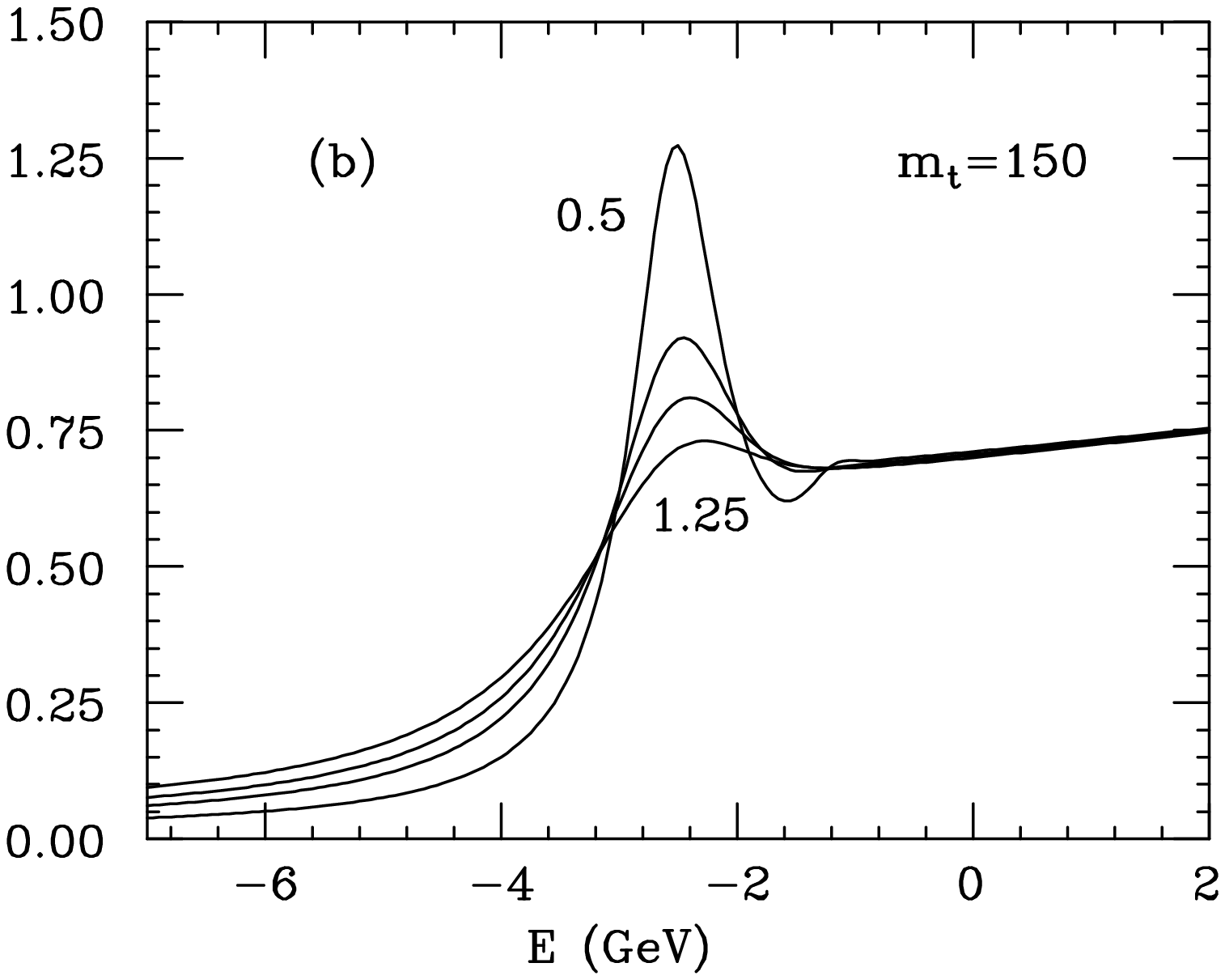}}
\vskip5pt
\vskip-2.7cm
\baselineskip=12pt
\centerline{\parbox{6.0in}{ Fig.~1:  The $t\bar t$ cross section
at threshold.  The curves in (a) correspond to
$\alpha_s=$ .10, .11, .12, and .13. The curves in (b) are with $\alpha_s$
fixed to .12 and the top width set to 0.5, 0.8, 1.0, and 1.25 times its
Standard Model value.}}
\vskip14pt
\end{figure}
In Fig.~1(a) I present a plot\cite{michael} of the
cross section as a function of energy for a top quark of 180 GeV
and several different values of $\alpha_s$.
One can see from this plot that the measurements
of $m_t$ and $\alpha_s$ are highly correlated; i.e., a change in $m_t$ can be
substantially mimicked by a change in $\alpha_s$.  It is still
possible, however, to get a very precise value of $m_t$.  In a detailed
Monte Carlo analysis using 11 cross section measurements of 1 fb$^{-1}$
each, and assuming a nominal top mass of 170 GeV, Fujii
{\it et al.}\cite{Fujii} were able to extract $m_t$ with an
estimated error of 380 GeV with $\alpha_s$ unconstrained.  Certainly,
the error will be even smaller if $\alpha_s$ is known precisely from
an independent experiment.

The effect on the cross section of a nonstandard top quark width
is shown in Fig.~1(b) for a top quark of 150 GeV.  Assuming
$\alpha_s$ is known exactly, Fujii {\it et al.} found that
the width could be measured with a relative error of 18\%.
If the standard model Higgs is light enough, it can also affect the
cross section by giving an additional contribution to the potential of
$\Delta V=(\lambda_t^2/4\pi)(1/r)\exp(-m_Hr)$.  For Higgs bosons of
less than about
100 GeV, Fujii {\it et al.} found that $\lambda_t$ should be
measurable to about 25\%.

Beyond simply measuring total $t\bar t$ cross section, one can probe the
same parameters with different error correlations by looking at kinematic
distributions of the top quark.  For example, the top quark momentum
distribution can be obtained from the Green's function solution to equation
(2) by\cite{kuhn}
\begin{equation}
{d\sigma\over d^3\vec p}\ \sim\ \Bigl|\tilde G_{E+i\Gamma_t}(\vec p)\Bigr|^2.
\end{equation}
Similarly, the top quark forward-backward asymmetry at threshold, arising from
the interference between the different $t\bar t$ angular momentum states will
be observable.

\section{The Top in Continuum}

In continuum $e^+e^-\rightarrow t\bar t$ production the obvious way to
measure $m_t$ is by kinematic reconstruction of the event, in a
manner similar to that used at the Tevatron.  The advantage of the NLC
is that the initial state is colorless, so that the events will be cleaner.
However, the treatment of QCD radiation in the final state must still
be considered.  The possibility of gluons radiating
off the top quarks and the bottom quarks, both before and
after the top decay, renders this a nontrivial problem.  In fact, recently
there have been reports of discrepancies between ${\cal O}(\alpha_s)$
calculations and standard Monte Carlo programs\cite{orr}.

\begin{figure}
\vskip10pt
\vskip-2.7cm
\epsfysize=11cm
\centerline{\hskip.5cm\epsffile{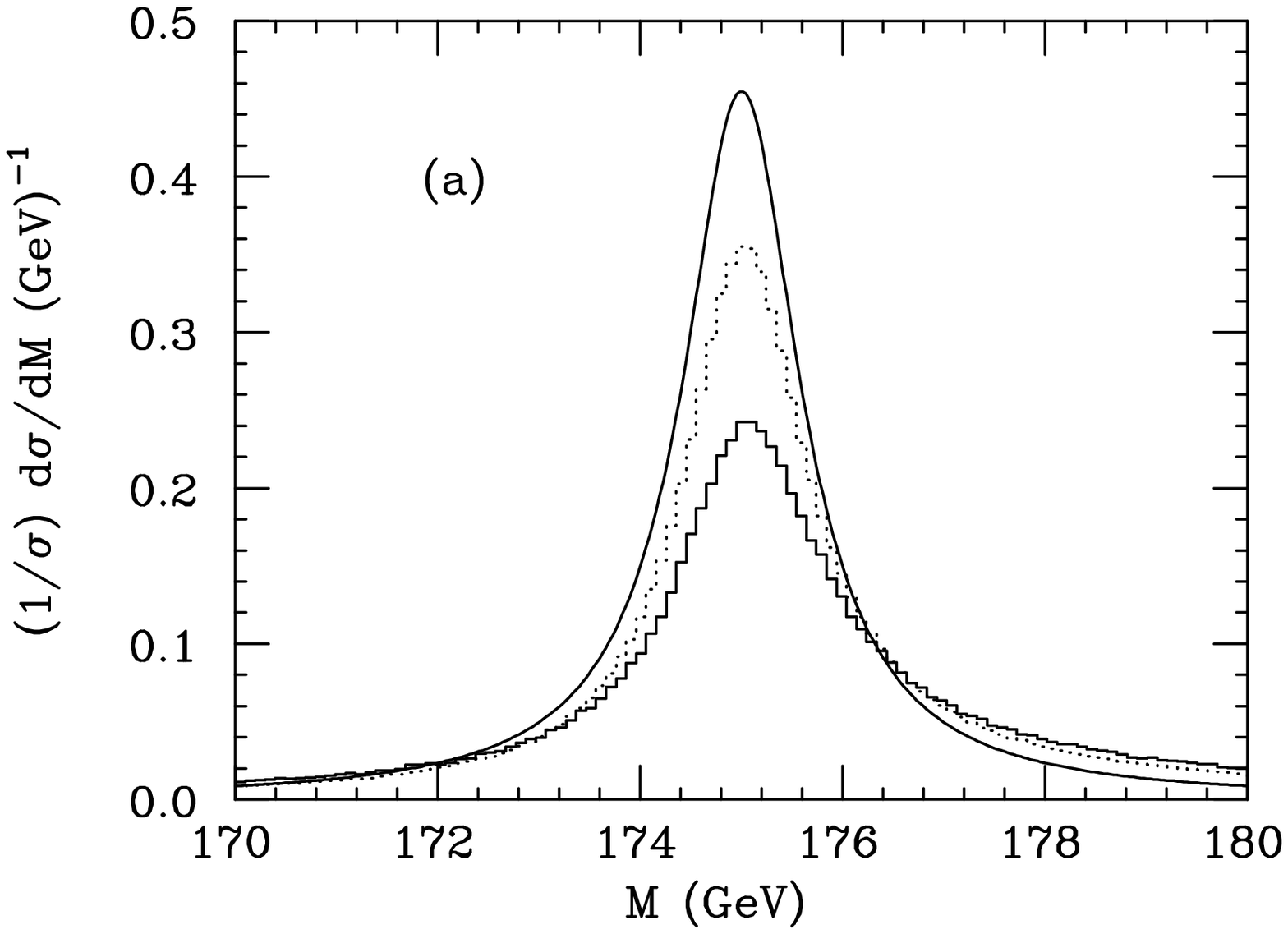}\hskip-1cm
            \epsfysize=11cm\epsffile{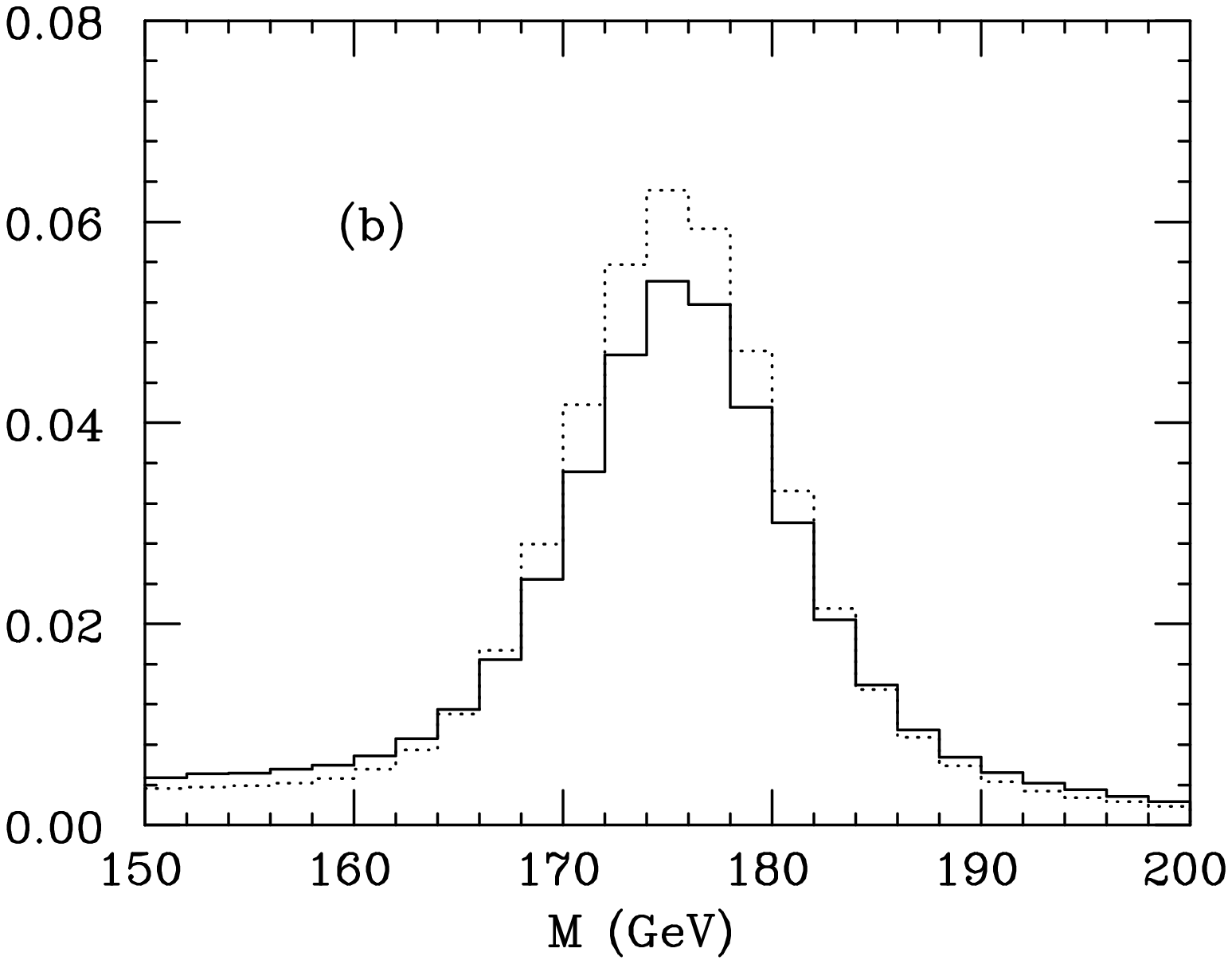}}
\vskip5pt
\vskip-2.7cm
\baselineskip=12pt
\centerline{\parbox{6.0in}{ Fig.~2: Top mass reconstruction
distributions in the lepton+jets mode
without (a) and with (b) detector energy smearing.
The dotted histograms are at tree-level, the solid histograms are
at ${\cal O}(\alpha_s)$, and the smooth curve in (a) is the original
Breit-Wigner distribution.}}
\vskip14pt
\end{figure}
Although a complete
phenomenological analysis of the continuum production is yet to be done,
I have performed a simplified analysis at ${\cal O}(\alpha_s)$ in the
production and the decay of the top quark, using the narrow top-width
approximation\cite{schmidt}.
Convoluting with a Breit-Wigner line shape for the top
quark squared-momentum produces an infrared-finite distribution in
perturbative QCD.  Fig.~2(a) displays the mass distribution for a 175 GeV
top quark in the lepton+jets mode at a 400 GeV center-of-mass collider.
The existence of an additional radiated gluon in the production or
decay of the top quark can confound the event reconstruction, producing
the degradation of the signal at ${\cal O}(\alpha_s)$.
Note, however, that the peak value of the distribution has not shifted.

In Fig.~2(b) I show the same distribution, but with Gaussian
smearing of the final-state lepton and jet energies due to the
detector resolution.  The smearing parameters used are
\begin{equation}
{\sigma_E^{had}\over E}\ =\ {0.4\over\sqrt{E}}\ ,\qquad
{\sigma_E^{lep}\over E}\ =\ {0.15\over\sqrt{E}}\ .
\end{equation}
{}From this we see that, although QCD radiation effects are certainly
non-negligible, the dominant systematic error will
probably be due to detector resolution.
The mass measurement in continuum is expected to be less
accurate than at threshold, but it will definitely be useful due to the
different systematic errors involved in the two determinations.

The process $e^+e^-\rightarrow t\bar t$ in continuum is
probably the best place
to study the couplings of the top quark to the photon and the weak gauge
bosons.
Expressed in terms of form
factors, the $\gamma,Z \rightarrow t\bar t $ production
vertices are
\begin{equation}
i{\cal M}^{i\mu} = ie
  \Bigl\lbrace
  \gamma^\mu  [Q^i_VF^i_{1V} + Q^i_AF^i_{1A}\gamma_5]
 +{i\sigma^{\mu\nu} q_\nu\over2m_t} [Q^i_VF^i_{2V} + Q^i_AF^i_{2A}
 \gamma_5]
  \Bigr\rbrace \ ,\label{prodffs}
\end{equation}
where the superscript is $i=\gamma,Z$.  In this formula
$Q^\gamma_V=Q^\gamma_A={2\over3}$,
$Q^Z_V=({1\over4}-{2\over3}s^2)/sc$, and $Q^Z_A=
(-{1\over4})/sc$, so that at tree level in the standard model
$F^\gamma_{1V}=F^Z_{1V}=F^Z_{1A}=1$ and all the others
form factors are zero.
The $t \rightarrow W^+b$ decay vertex can be treated in an
analogous fashion.
The form factors are typically corrected by
a few percent or less from QCD and electroweak loops.
Thus, any large deviations would
indicate new physics.

\begin{figure}
\vskip10pt
\vskip-2.7cm
\epsfysize=11cm
\centerline{\epsffile{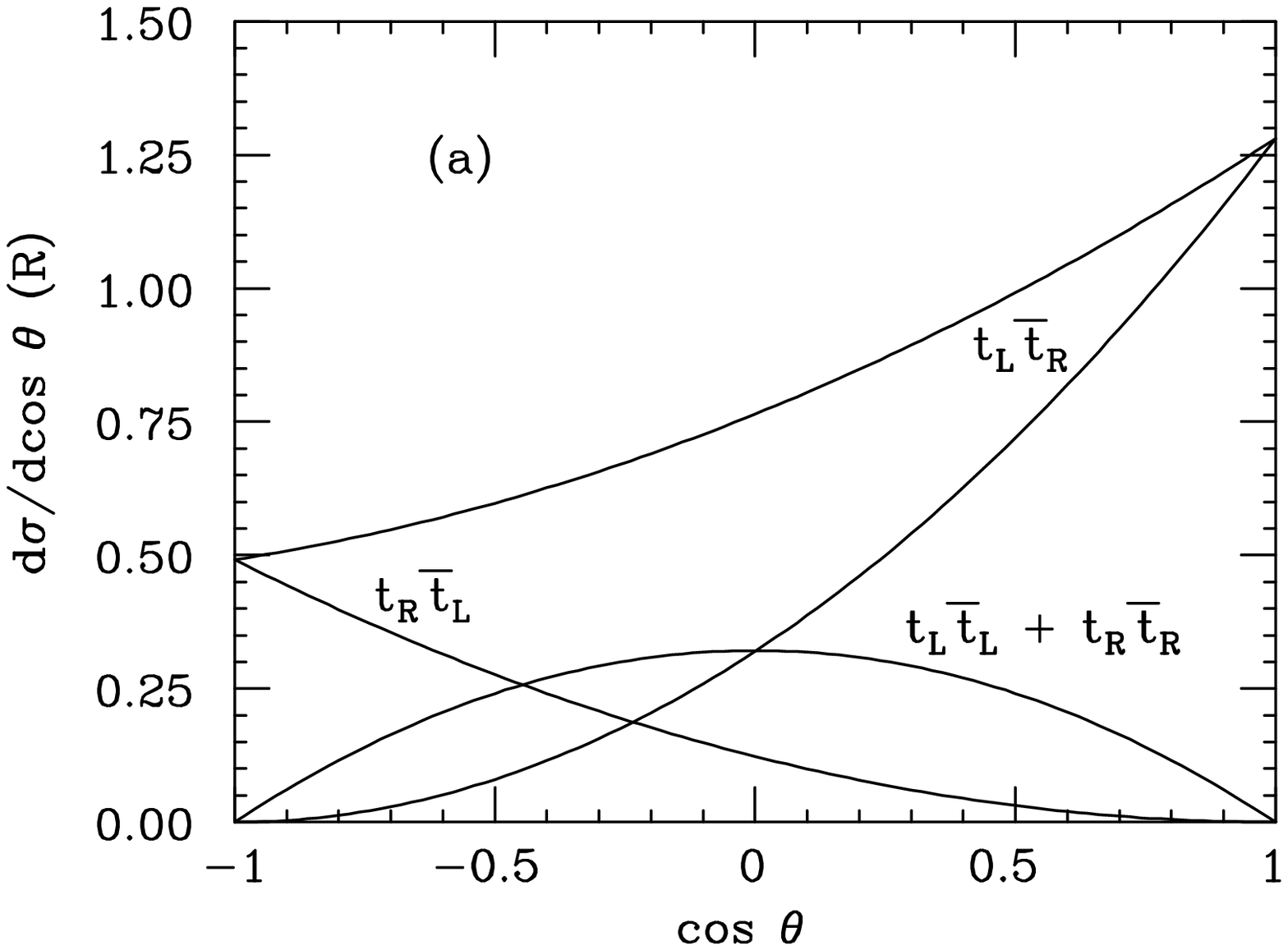}}
\vskip5pt
\vskip-2.7cm
\baselineskip=12pt
\centerline{ Fig.~3:  $t\bar t$ cross section for left-polarized
electrons.}
\vskip14pt
\end{figure}
Before beginning the analysis of the sensitivity of the NLC to these
couplings, it is useful to re-emphasize the importance of polarization
in the top quark interactions.  In Fig.~3 I show the
$t\bar t$ cross section
as a function of the top quark production angle for initial left-handed
polarized electrons.  By breaking down the event into helicity subprocesses,
we see that much of the electron polarization is passed on to the top
quarks.  This spin transfer is perfect in the forward direction so that
for large values of $\cos\theta$ we obtain a sample of highly polarized top
quarks.

To make the most of this polarization, we must consider all of the
helicity angles involved in the event.
With Tim Barklow, I performed a tree-level study of the sensitivity to a
maximum-likelihood analysis using all the information in the $t\bar t$
event\cite{barklow}.  The top mass was set to $m_t=174$ GeV, and the
the NLC parameters were chosen to be
$\sqrt{s}=400$ GeV, an integrated
luminosity of 100 fb$^{-1}$, and 90\% polarization for the electrons.
Only the lepton+jets mode was included in the analysis, and a simple
angular cut of $|\cos\theta_{\rm lab}|<0.8$ was applied to all of the visible
final-state particles.

\begin{figure}
\vskip10pt
\vskip-2.7cm
\epsfysize=11cm
\centerline{\epsffile{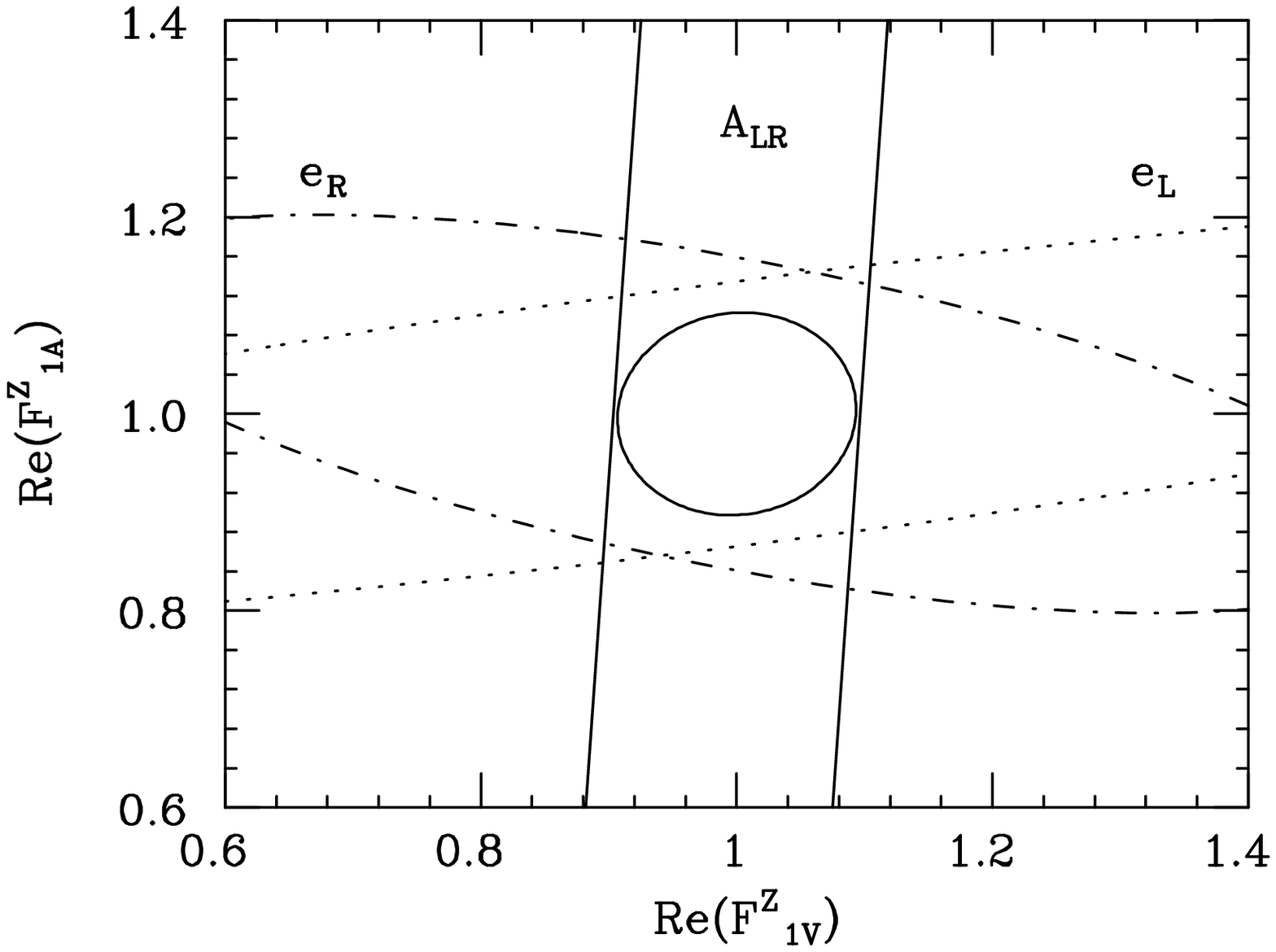}}
\vskip5pt
\vskip-2.7cm
\baselineskip=12pt
\centerline{Fig.~4:  95\% confidence level contours.}
\vskip14pt
\end{figure}
In Fig.~4 I plot the 95\% confidence level contours
for $F_{1V}^Z$ and $F_{1A}^Z$, obtained from the maximum-likelihood
analysis using a sample of 50 fb$^{-1}$ each of right- and
left-polarized electrons.  This exhibits the degree to which we can
 probe the anomalous couplings of the top quark to the $Z$ boson
at the NLC.
Note that the measurement of $A_{LR}=(\sigma_L-\sigma_R)/
(\sigma_L+\sigma_R)$ is crucial for constraining the vector coupling
to the $Z$ boson.  In fact, the use of polarized beams increases
the sensitivity to this coupling by as much as a factor of 2.
The combined 95\% confidence level for the vector and axial-vector
couplings yields an error
of about 10\% for this luminosity and these cuts.  Surprisingly, the
full maximum-likelihood analysis does not provide much extra
sensitivity over an analysis based solely on the top quark production
angle distribution---as long as the electron beam is polarized.
A phenomenological analysis by Ladinsky and Yuan\cite{ladyuan}
of the production angle distribution of the top quark
is consistent with these results.

In summary, there is much beautiful physics of the top quark to be
explored at the Next Linear Collider.  In this
talk I have only touched the surface.

\vglue 0.5cm

\vglue 0.5cm


\begin{thebibliography}{99}
\bibitem{michael}
I would like to thank Michael Peskin for supplying me with Figs.~1(a) and (b).
\end{thebibliography}
\end{document}